\documentclass[11pt,twoside]{article}
\usepackage{asp2010}
\resetcounters

\begin{document}

\title{SLUG:  A new way to Stochastically Light Up Galaxies}
\author{Michele Fumagalli$^1$, Robert da Silva$^{1\dag}$, Mark Krumholz$^1$, and Frank Bigiel$^2$
\affil{$^1$ Department of Astronomy and Astrophysics, University of California,  
  1156 High Street, Santa Cruz, CA 95064}
\affil{$^2$ Department of Astronomy and Astrophysics, University of California,  
  601 Campbell Hall, Berkeley, CA 94720}
\affil{$^\dag$ NSF graduate research fellow}}

\begin{abstract}
We present {\scshape SLUG}, a new code to ``Stochastically Light Up Galaxies''.
{\scshape SLUG} populates star clusters by randomly drawing 
stars from an initial mass function (IMF) and then following their time evolution with stellar 
models and an observationally-motivated prescription for cluster disruption. 
For a choice of star formation history, metallicity, and IMF, 
{\scshape SLUG} outputs synthetic photometry  for  clusters and field stars 
with a proper treatment of stochastic star formation. {\scshape SLUG} generates 
realistic distributions of star clusters, demonstrating the range of properties that result from
finite sampling of an IMF and a random distribution of ages.
The simulated data sets provide a quantitative means to address open problems in 
studies of star formation in galaxies and clusters, such as 
a test for IMF variations that are suggested by the systematic deficiency in the 
H$\alpha$/UV ratio in outer disks or in dwarf galaxies.
{\scshape SLUG} will be made publicly available through the website 
{\tt http://sites.google.com/site/runslug/}.

\end{abstract}

\section{Motivations}
The continuous build up of observations is enriching our view of how galaxies form and evolve, 
but not without introducing new riddles. The availability of \emph{Galex} 
UV data, together with ground-based H$\alpha$ imaging, 
has recently uncovered a systematic deficiency of hydrogen recombination line emission 
normalized to UV fluxes below star formation rates (SFRs) 
of $\lesssim 0.1$ M$_{\odot}$ yr$^{-1}$. 
A debate around the origin of this H$\alpha$ deficit 
has opened, questioning some of the basic assumptions in star formation studies, 
such as a constant initial mass function (IMF) \citep[e.g.][]{meu09,lee09}. 

Support to this interpretation comes from the apparent 
inability of spectral energy distribution (SED) modeling with different star formation 
histories (SFHs) \citep[e.g.][]{hov08} 
or stochastic sampling of canonical IMFs with a constant SFR \citep[e.g.][]{lee09}
to fully account for the observed deficiency of H$\alpha$. Furthermore, semi-empirical 
models of a galactic IMF that depend on the SFR
appear to reproduce the observed trends \citep[e.g. the IGIMF theory;][]{pfl09}. 
However, uncertainties on
dust corrections, stellar models, star formation histories, 
or escape fraction of ionizing radiation 
limit our ability to unambiguously interpret these observations 
\citep[e.g.][but see Meurer et al.~2009]{bos09}.

Beside the intrinsic difficulties of understanding this effect,  
a straightforward comparison of independent studies is made even harder 
by the inhomogeneity in the various data sets and in the stellar models 
used for theoretical predictions. 
To help addressing the fundamental problem 
of the IMF variation or more generally other open issues in studies of star formation,
we present {\scshape SLUG}, a new code to ``Stochastically Light Up Galaxies''.
Among its applications, {\scshape SLUG} can be used to extensively test the null hypothesis 
of incomplete sampling of the upper end of the IMF for low star formation rates as well 
as to the test for environmental effects (e.g., metallicity) on the observed H$\alpha$/UV ratio.

\section{The SLUG code}\label{code}

\begin{figure}
\plotone{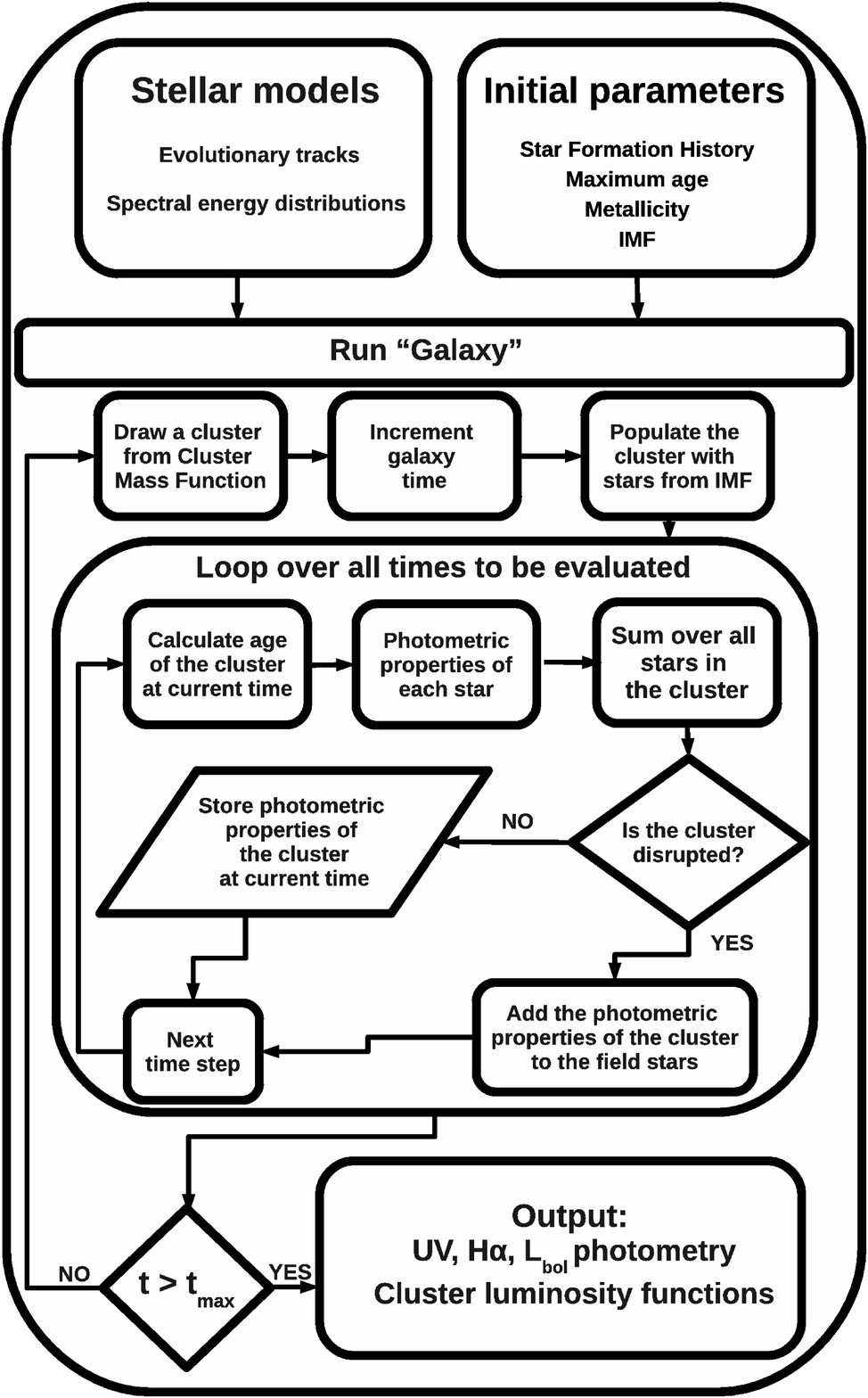}
\caption{Schematic representation of the {\scshape SLUG} code. 
See Section \ref{code} for a description. \label{fig:slugflow}}
\end{figure}

A schematic representation of the {\scshape SLUG} code is shown in Figure \ref{fig:slugflow}.
A more extensive description will be presented in 
da Silva et al. (in preparation). As inputs, {\scshape SLUG} accepts an arbitrary SFH, metallicity, 
and a choice of IMF (Chabrier, Kroupa, Salpeter or IGIMF). From these parameters, 
the code creates a collection of star clusters and field stars, representative of 
a portion of a galaxy. A more realistic galaxy can then be assembled by 
combining different outputs tuned to best match the SFH and metallicity properties 
across the objects of interest.

In more detail, the code starts by drawing a cluster mass from a given cluster mass function 
of the form $dN/dM\propto M^{-\beta}$, extending from $M_{\rm cl,min}$ to $M_{\rm cl,max}$ 
(by default $\beta=2$, $M_{\rm cl,min}=20 M_\odot$, and $M_{\rm cl,max}=10^7 M_\odot$). 
The birth of a cluster represents a new event in the galaxy SFH at time $t$ and 
therefore the age of the galaxy is incremented as implicitly defined by  
\begin{equation}\label{age}
M_{\rm galaxy}(t)=\int_0^t SFR(t')dt'\:.
\end{equation}
The next step is to populate the cluster with stars randomly 
drawn from the chosen IMF. The combination of these two random draws
(for clusters and stars) allows for a proper treatment of stochastic
star formation.

Once the cluster has been filled, the photometric properties  (far UV and near UV in the \emph{Galex}
passbands, bolometric luminosity, and H$\alpha$) of individual stars are computed for a series of 
time steps, up to a maximum age defined by the user. 
To this purpose, {\scshape SLUG} uses libraries of theoretical 
evolutionary tracks \citep{sch92} combined with model spectral energy distributions 
\citep{lej01,gir04}\footnote{Additional stellar 
models will be provided in the public version of the code or can be 
added by the user.}. Next, the  code outputs the integrated cluster photometry by 
summing over all the star luminosities in each cluster. 

Clusters dissolve after $\sim 1$ Myr from their formation 
at a rate $\propto t^{-1}$ \citep{fal09} and
{\scshape SLUG} reproduces this empirical disruption rate by evaluating at each time 
step the probability that a cluster survives.
If the cluster is disrupted, the code continues to follow the evolution of the individual 
stars, but the stellar integrated luminosities are added to a field where all the former 
cluster-member stars are accumulated.

These operations are repeated iteratively for the subsequent star clusters, 
until the age of the galaxy as defined by equation (\ref{age}) reaches the maximum age 
specified by the user.
As a last step, {\scshape SLUG} writes a series of binary files containing
the mass, age, photometry, number of stars
and the maximum stellar mass for each cluster at each time step.
Also, the photometric properties of all the stars from disrupted clusters 
are recorded in histograms. 
With a suite of {\scshape IDL} codes\footnote{The IDL procedures will be distributed 
as part of {\scshape SLUG}.}, 
useful statistics such as the cluster luminosity functions or probability distribution
functions can easily be generated.

\section{Applications and future developments}

An immediate application of {\scshape SLUG} is to verify whether
stochastic star formation may explain the observed decrease in the UV/H$\alpha$ ratio 
at low SFRs. Indeed, incomplete sampling of the 
IMF combined with stellar evolution and cluster disruption produces 
populations that are deficient in high mass stars, even if the underlying stellar IMF 
is universal and independent of the star formation rate. 
{\scshape SLUG} will enable an extensive and homogeneous comparison between observations and 
theoretical predictions over a large parameter space (SFHs, metallicity, IMFs, ages, stellar models). 
Particularly useful will be a comparison of outputs from various stellar models
to quantify uncertainties in theoretical predictions, particularly relevant to 
test modest variation in the IMF.

{\scshape SLUG} can also be used to generate synthetic distributions for
the cluster mass and the maximum mass of cluster-member stars \citep[e.g.][]{wei06}, 
or to test and develop new analysis techniques which are based on 
probabilistic formulations \citep[e.g.][]{cer06}, or more generally for 
applications related to photometric distributions of clusters. 
In future versions of the code, we aim to output additional broad-band photometry and 
spectral properties of clusters and field stars, 
suitable to simulate color-magnitude diagrams from random sampling of the IMF. 

{\scshape SLUG} will be publicly available at {\tt http://sites.google.com/site/runslug/}.
It is easy to use and modify, and users can include physical processes 
relevant for specific applications.

\acknowledgements  {\small 
We would like to thank the UP2010 LOC and SOC for 
organizing a very nice conference in a wonderful setting. 
MRK acknowledges support from: an Alfred P. Sloan Fellowship; 
NASA through ATFP grant NNX09AK31G; NASA as part of the Spitzer Theoretical 
Research Program, through a contract issued by the JPL; the National Science Foundation through grant AST-0807739. The work of RdS is supported by a National Science 
Foundation Graduate Research Fellowship. MF is supported by NSF grant (AST-0709235).}

\bibliography{fumagalli_m}

\end{document}